\documentclass[useAMS,usenatbib]{mn2e}
\usepackage{graphicx}
\usepackage{lscape}
\usepackage{multirow}
\usepackage{slashbox}
\usepackage{amsmath,amssymb}
\usepackage{natbib}
\usepackage{color}
\title[Infared galaxies at z$\sim$1]{Extinction law in ultraluminous infrared galaxies at z$\sim$1}
\author[T. Shimizu et al.]{T. Shimizu$^{1}$\thanks{E-mail:
tshimizu@ioa.s.u-tokyo.ac.jp}, K. Kawara$^{1}$\thanks{E-mail:
kkawara@ioa.s.u-tokyo.ac.jp}, H. Sameshima$^{1}$, N. Ienaka$^{1}$,
T. Nozawa$^{2}$, and T. Kozasa$^{3}$\\
$^{1}$Institute of Astronomy, the University of Tokyo, Osawa 2-21-1, Mitaka, 
Tokyo 181-0015, Japan\\
$^{2}$Institute for the Physics and Mathematics of the Universe, 
the University of Tokyo,
 5-1-5 Kashiwanoha, Kashiwa, 277-8568, Japan\\
$^{3}$Department of Cosmosciences, Hokkaido University, 
Sapporo 060-0810, Japan}
\begin{document}

\date{MNRAS in press: Accepted 2011 July 26. Received 2011 July 26; in original form 2011 May 05}

\pagerange{\pageref{firstpage}--\pageref{lastpage}} \pubyear{2011}

\maketitle

\label{firstpage}

\begin{abstract}

We analyze the multi-wavelength photometric and spectroscopic data of 12 ultraluminous infrared 
galaxies (ULIRGs) at $z \sim 1$ and compare them with models of stars and dust in order 
to study the extinction law and star formation in young infrared (IR) galaxies.
Five extinction curves, namely, the Milky Way (MW), the pseudo MW which is MW-like 
without the 2175 \AA\ feature, the Calzetti, and two SN dust curves, are applied to the data, 
by combining with various dust distributions, namely, the uniform dust screen, 
the clumpy dust screen, the internal dust geometry, and the composite geometry with a combination of 
dust screen and internal dust. 
Employing a minimum  $\chi^2$ method, we find that the foreground dust screen geometry, 
especially combined with the 8 - 40 $M_{\sun}$ SN extinction curve, 
provides a good approximation to the real dust geometry, whereas 
internal dust is only significant in 2 galaxies. 
The SN extinction curves, which are flatter than the others, reproduce 
the data of 8(67\%) galaxies better. 
Dust masses are estimated to be in excess of 
$\sim 10^8 M_{\sun}$. Inferred ages of the galaxies are very young, 8 of 
which range from 10 to 650 Myr. 
The SN-origin dust is the most plausible to account for the vast amount 
of dust masses and the flat slope of the observed extinction law.
The inferred dust mass per SN ranges from 0.01 to 0.4 M$_{\sun}$/SN. 

\end{abstract}


\section{Introduction}

Recent developments in far-infrared(IR) and submillimeter astronomy have revealed 
new populations of galaxies associated with thermal dust emission, such as dusty quasars 
and infrared galaxies. Thermal dust emission detected 
in quasars and galaxies at high redshift exhibits inferred dust mass in excess of 
$\sim 10^{8} M_{\sun}$. Such a vast amount of dust observed at $z \gtrsim 6$ 
(\citealt{Bertoldi2003}; \citealt{Robson2004};\citealt{Beelen2006}) strongly suggests that 
the dust has been condensed in ejecta of core-collapsed supernovae (SNe II) promptly 
after the onset of the initial star formation. 
Asymptotic giant branch (AGB) stars are known to be the major source of dust in 
the present-day universe. However, the importance of AGB stars at high redshift is in controversy, 
because they inject dust grains into the ISM only after their progenitor 
stars have evolved off the main sequence. While a delay time of $\gtrsim$ 500 Myr 
since their formation has been suggested (e.g., \citealt{Dwek1998}), \citet{Valiante2009} 
proposed that AGB stars can dominate dust production in a delay time 
of $\gtrsim$ 150 Myr if the starburst is instantaneous.
\citet{Michalowski2010} pointed out that AGB stars are not efficient enough to form dust in dusty 
quasars at $z > 5$, whereas SNe II may be able to account for the dust, and underlined 
the importance of dust grain growth in the interstellar medium (ISM).  
 
Quasars have been frequently used to study the extinction law
at high redshift. This is because of their simple dust geometry,
in which the central point-like emitting source is surrounded by
dust grains which are physically separated from the source itself.
This is analogous to the classical configuration assumed when
dereddening single stars.
There is a line of evidence that active galactic nuclei
(AGNs) have extinction curves that differ from those observed in the Milky Way (MW), 
Large and Small Magellanic Clouds (LMC and SMC). \citet{Gaskell2004} analyzed a large sample 
of AGNs and found indications of a very flat extinction curve in the ultra-violet(UV). 
On the other hand, exploring large samples of quasars from the Sloan Digital Sky Survey(SDSS), 
\citet{Richards2003} and \citet{Hopkins2004} found that the extinction law in quasars 
at $z < 2.2$ is described by SMC reddening but neither by LMC and SMC reddening nor by 
that found in the \citet{Gaskell2004} sample. Recently, \citet{Gallerani2010} proposed 
that quasars at $z = 3.9 - 6.4$ have reddening that deviates from that of the SMC with 
a tendency flatten at $\lambda \lesssim$ 2000 \AA. \citet{Maiolino2004} found that 
SDSS J1048+4647, a broad absorption line quasar (BALQSO) at $z$=6.2,
requires a flat extinction curve which is quite different from those observed in the MW, LMC 
and SMC, but in excellent agreement with SN extinction curves by \citet{Todini2001}.

Although it is reasonable to expect that the extinction law in young galaxies 
would be dominated by SN-condensed dust, the complications introduced by dust geometries 
have hampered this line of studies; the attenuation of extended starlight by dust depends not only 
on the optical properties of dust but also on the distribution geometry 
of the dust. \citet{Calzetti2001} derived the empirical extinction law in local starbursts 
and blue compact galaxies, assuming the dust geometry described in a foreground screen, 
i.e., an apparent attenuation of starlight to be $e^{-\tau_{\nu}}$. 
If a dust geometry in their galaxies is indeed a foreground screen, 
the slope (i.e.,wavelength dependence) of the extinction curve (i.e., 
the optical depth $\tau_{\nu}$) is significantly flatter than those observed in the MW, LMC, and SMC.
The SED and spectroscopic data of a young, ultraluminous infrared galaxy (ULIRG)\footnote{
Infrared galaxies are classified according to bolometric infrared luminosity 
$L_{IR}(8 - 1,000\micron)$; $L_{IR} > 10^{11} L_{\sun}$ as luminous infrared galaxies (LIRGs), 
$L_{IR} > 10^{12}L_\odot$ as ultraluminous infrared galaxies (ULIRGs),and 
$L_{IR} > 10^{13}L_\odot$ as hyperluminous infrared galaxies (HyLIRGs) \citep{Sanders1996}.} 
at $z \sim 1$ were analysed by \citet{Kawara2011}, who found that the data require a flat 
extinction curve and a dust geometry of a foreground dust screen, resulting in better 
agreement with the extinction curve synthesized from SN-condensed dust by 
\citet{Hirashita2005} than those observed in local starbursts, MW, LMC, and SMC.

The slope of the extinction curve is expected to reflect the sites of dust formation 
and the subsequent processing in the ISM. The extinction curves can be ordered  as
the SMC, LMC, MW, and Calzetti curves from the steepest to the flattest. 
The synthetic extinction curves of newly condensed dust grains in SN ejecta before the ISM 
reprocessing (hereafter called original SN extinction curves) are flatter than the Calzetti 
curve (\citealt{Todini2001}; \citealt{Nozawa2003}; \citealt{Maiolino2004}; \citealt{Hirashita2005}; 
see also Figure 3 in \citealt{Kawara2011}). Then, the bulk of dust grains condensed in the ejecta 
is destroyed through the passage of the reverse shock 
(\citealt{Bianchi2007};\citealt{Nozawa2007};\citealt{Hirashita2008}). Finally, 
grains that survive the reverses shock are shattered in the warm ionized ISM \citep{Hirashita2010}. 
Nonetheless, the final SN extinction curves, after taking into account the reverse shock and 
shattering, are expected to have a slope as flat as those of the original extinction curves. 
The shattering may also be important in old galaxies like MW, LMC, and SMC, 
where AGB stars are dominant sources of dust production, because some observations 
(\citealt{Groenewegen1997};\citealt{Gauger1999}) suggest AGB stars form large grains.
Dust grain growth in the ISM would be efficient in infrared galaxies where there must be 
many molecular clouds near SN remnants, and it is suggested that the dust mass
in the high-redshift quasars is dominated by grain growth in the IMS 
(\citealt{Dwek1998};\citealt{Draine2009};\citealt{Michalowski2010};\citealt{Valiante2011}). 
It seems that this process result in flat extinction curves in the sense that the larger grains 
show the flatter extinction curve. However, the extinction laws
found in the high-redshift quasars and local starbursts are flatter than those found in the 
MW for which \citet{Draine2009} attributes the bulk of the dust mass to the growth of grains 
in the ISM.

As discussed so far, $z < 2.2$ quasars show a striking contrast in the slopes of the extinction 
curves against the $z \sim 1$ ULIRG and higher redshift quasars. 
In this paper, we explore this issue by expanding our previous work 
on a single young ULIRG \citep{Kawara2011} to 12 young ULIRGs at $z \sim 1$.

\section{Data}
\begin{table}
	\begin{minipage}{85mm}
		\centering
		\caption{Sample galaxies at $z \sim 1$}
		\begin{tabular}{cccc}
		\hline
			No. & Name & Redshift & Reference \\
		\hline
			1 & COSMOS J100121.29+023535.8 & 0.850 & 1 \\
			2 & COSMOS J100125.12+024506.5 & 0.980 & 1 \\
			3 & COSMOS J100158.31+023505.2 & 1.021 & 1 \\
			4 & COSMOS J100143.83+015325.0 & 1.066 & 1 \\
			5 & COSMOS J100021.79+014916.1 & 0.995 & 1 \\
			6 & COSMOS J100030.12+020513.3 & 0.894 & 1 \\
			7 & COSMOS J100123.22+021013.3 & 0.850 & 1 \\
			8 & COSMOS J100100.22+021503.6 & 0.893 & 1 \\
			9 & COSMOS J100008.36+022244.6 & 1.093 & 1 \\
		       10 & COSMOS J095910.21+022523.3 & 1.026 & 1 \\
		       11 & SST J1604+4304 & 1.135 & 2 \\
		       12 & HR10 (ERO J164502+4626.4) & 1.44 & 3, 4\\
		\hline
		\end{tabular}
		\label{t_sample}
		\begin{flushleft}
		\medskip
\textbf{Reference.} - 
(1) Cosmos archive at http://irsa.ipac.caltech.edu: \\
COSMOS Photometry Catalog January 2006:\\
S-COSMOS IRAC 4-channel Photometry Catalog June 2007;\\ 
S-COSMOS MIPS 24$\mu $m Photometry Catalog  October 2008; \\
S-COSMOS MIPS 70$\mu $m Photometry Catalog v3 January 2009; \\
COSMOS MAGELLAN Data Version 3.1 (spectroscopy);\\
zCOSMOS data released in 2008 October (spectroscopy); \\
(2) \citealt{Kawara2010}; (3) \citealt{Stern2006}; (4) \citealt{Dey1999}.
		\end{flushleft}
	\end{minipage}
\end{table}

\begin{figure*}
	\begin{minipage}{170mm}
		\centering
		\begin{tabular}{cl}
		\hline
No. &
\hspace{0pt} F814W \hspace{10pt} u$^*$ \hspace{16pt} g$+$ \hspace{16pt} r$+$ \hspace{18pt} i$+$ \hspace{18pt} z$+$
\hspace{16pt} Ks \hspace{12pt} 3.6\micron \hspace{7pt} 4.5\micron \hspace{7pt} 5.8\micron \hspace{7pt} 8.0\micron
\hspace{7pt} 24\micron \hspace{10pt} 70\micron \\
		\hline
		\hline
			\multirow{-4}{*}{1}  & \includegraphics[width=15cm,clip]{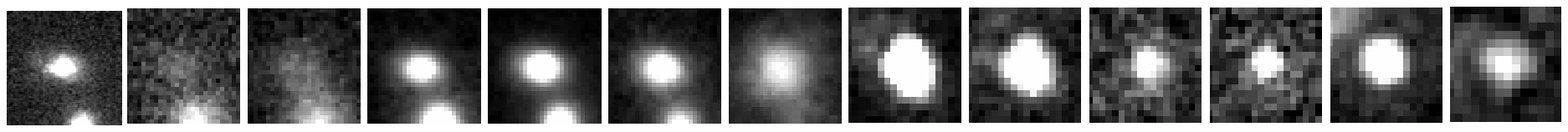} \\
			\multirow{-4}{*}{2}  & \includegraphics[width=15cm,clip]{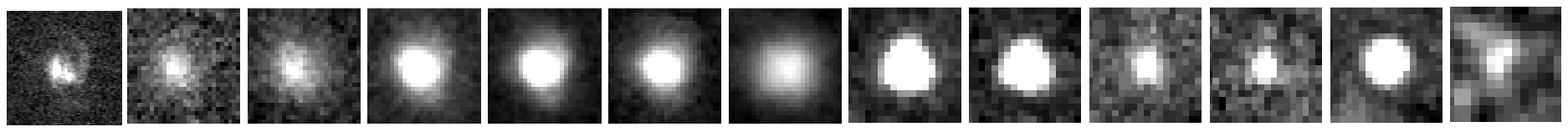} \\
			\multirow{-4}{*}{3}  & \includegraphics[width=15cm,clip]{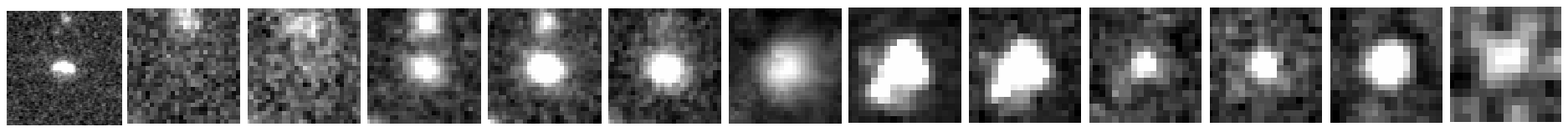} \\
			\multirow{-4}{*}{4}  & \includegraphics[width=15cm,clip]{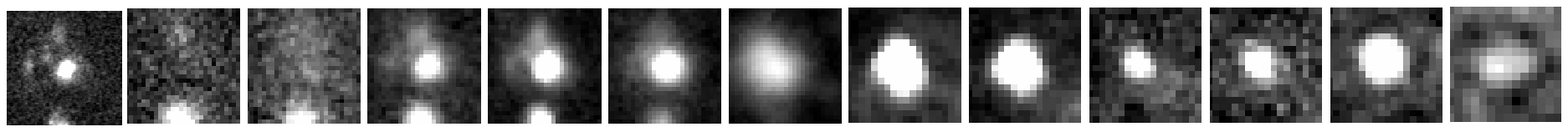} \\
			\multirow{-4}{*}{5}  & \includegraphics[width=15cm,clip]{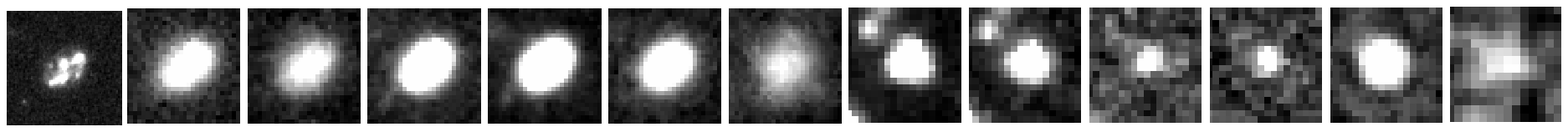} \\
			\multirow{-4}{*}{6}  & \includegraphics[width=15cm,clip]{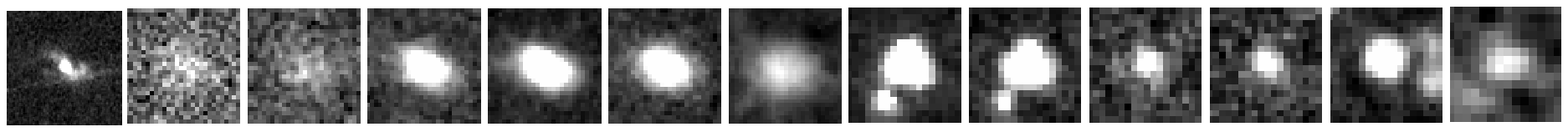} \\
			\multirow{-4}{*}{7}  & \includegraphics[width=15cm,clip]{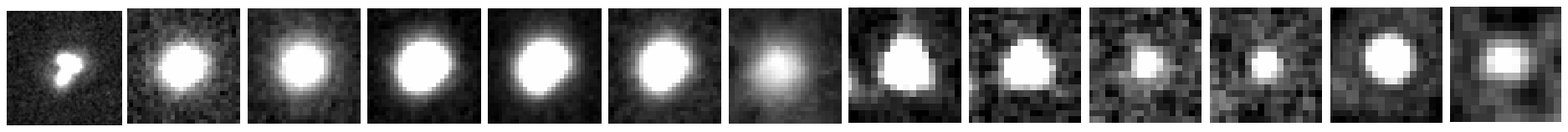} \\
			\multirow{-4}{*}{8}  & \includegraphics[width=15cm,clip]{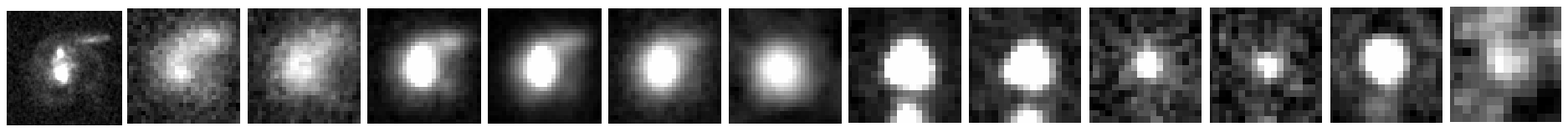} \\
			\multirow{-4}{*}{9}  & \includegraphics[width=15cm,clip]{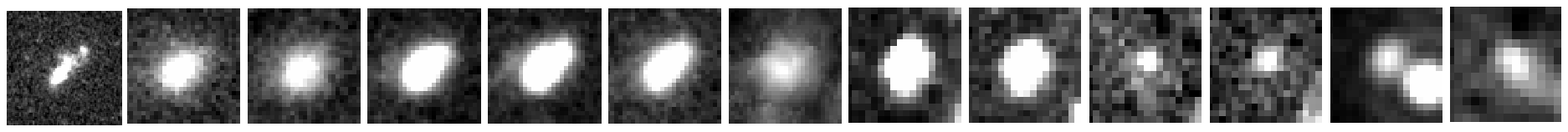} \\
			\multirow{-4}{*}{10} & \includegraphics[width=15cm,clip]{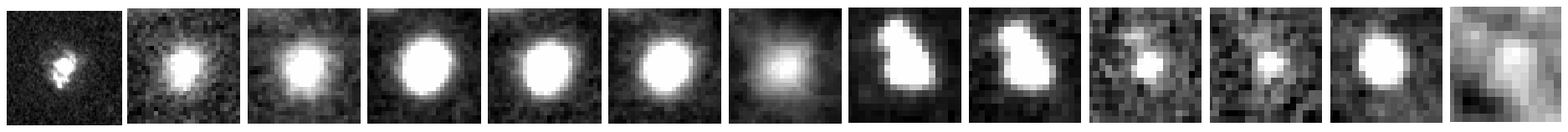} \\
\hline
\end{tabular}
\caption{Postage stamp images of the 10 COSMOS galaxies.
The sizes of each images are 4\arcsec$\times$4\arcsec\ for F814W to Ks, 
10\arcsec$\times$10\arcsec\ for 3.6\micron to 8.0\micron, 
20\arcsec$\times$20\arcsec\ for 24\micron, and 50\arcsec$\times$50\arcsec\ for 70\micron.
North is up and east is to the left.
Refer to \citet{Kawara2010} and \citet{Stern2006} for the images of No.11 (SST J1604+4304) and No.12 (HR10), 
respectively.}
		\label{f_photo}
	\end{minipage}
\end{figure*}

\begin{table*}
	\begin{minipage}{150mm}
		\centering
		\caption{Optical and 2\micron\ photometry of the 10 COSMOS galaxies}
		\begin{tabular}{ccccccc}
		\hline
     No. & u$^*$ & g$+$ & r$+$ & i$+$ & z$+$ & Ks \\
		\hline
1 & 0.530 $\pm$ 0.033 & 0.924 $\pm$ 0.041 & 3.04 $\pm$ 0.07 & 8.43 $\pm$ 0.10 & 14.2 $\pm$ 0.2 & 62.6 $\pm$ 3.0 \\
2 & 0.888 $\pm$ 0.048 & 1.27 $\pm$ 0.05 & 3.05 $\pm$ 0.07 & 7.76 $\pm$ 0.11 & 13.5 $\pm$ 0.2 & 63.8 $\pm$ 2.9 \\
3 & 0.096 $\pm$ 0.022 & 0.164 $\pm$ 0.024 & 0.656 $\pm$ 0.034 & 1.95 $\pm$ 0.06 & 3.78 $\pm$ 0.11 & 38.0 $\pm$ 3.0 \\
4 & 0.892 $\pm$ 0.052 & 1.50 $\pm$ 0.07 & 3.97 $\pm$ 0.09 & 9.52 $\pm$ 0.14 & 19.2 $\pm$ 0.2 & 97.8 $\pm$ 4.6 \\
5 & 2.20 $\pm$ 0.06   & 2.51 $\pm$ 0.06 & 5.10 $\pm$ 0.08 & 9.46 $\pm$ 0.11 & 13.8 $\pm$ 0.2 & 44.0 $\pm$ 3.4 \\
6 & 0.225 $\pm$ 0.023 & 0.403 $\pm$ 0.003 & 1.38 $\pm$ 0.05 & 4.43 $\pm$ 0.08 & 7.91 $\pm$ 0.13 & 47.6 $\pm$ 2.9 \\
7 & 2.07 $\pm$ 0.06   & 3.22 $\pm$ 0.07 & 6.50 $\pm$ 0.09 & 12.6 $\pm$ 0.1 & 18.4 $\pm$ 0.2 & 49.3 $\pm$ 3.1 \\
8 & 1.23 $\pm$ 0.05   & 1.97 $\pm$ 0.06 & 4.60 $\pm$ 0.09 & 11.1 $\pm$ 0.1 & 18.1 $\pm$ 0.2 & 75.7 $\pm$ 3.5 \\
9 & 1.10 $\pm$ 0.04   & 1.26 $\pm$ 0.04 & 2.13 $\pm$ 0.06 & 4.03 $\pm$ 0.07 & 6.98 $\pm$ 0.12 & 30.8 $\pm$ 3.1 \\
10 & 1.13 $\pm$ 0.04  & 1.83 $\pm$ 0.05 & 3.68 $\pm$ 0.07 & 7.73 $\pm$ 0.10 & 13.8 $\pm$ 0.2 & 50.6 $\pm$ 3.0 \\
		\hline
		\end{tabular}
		\label{t_flux1}
		\begin{flushleft}
		\medskip
\textbf{Note.} - Total fluxes are given in units of $\mu$Jy.
		\end{flushleft}
	\end{minipage}
\end{table*}

\begin{table*}
	\begin{minipage}{170mm}
		\centering
		\caption{{\it Spitzer} photometry of the 10 COSMOS galaxies}
		\begin{tabular}{ccccccc}
		\hline		
 			No. & 
			3.6\micron & 4.5\micron & 5.8\micron & 8.0\micron & 
			24\micron & 70\micron \\
		\hline
			1 & 
			106 $\pm$ 0.2 & 81.6 $\pm$ 0.3 & 74.7 $\pm$ 1.1 & 76.5 $\pm$ 2.8 & 
			940 $\pm$ 252 & 20900 $\pm$ 3300 \\
			2 & 
			89.6 $\pm$ 0.2 & 73.1 $\pm$ 0.3 & 66.0 $\pm$ 1.2 & 58.9 $\pm$ 2.9 & 
			797 $\pm$ 19 & 11700 $\pm$ 2600 \\
			3 & 
			81.7 $\pm$ 0.2 & 66.5 $\pm$ 0.3 & 72.9 $\pm$ 1.2 & 82.1 $\pm$ 3.0 & 
			635 $\pm$ 20 & 11100 $\pm$ 2600 \\
			4 & 
			128 $\pm$ 0.7 & 111 $\pm$ 0.7 & 78.6 $\pm$ 1.3 & 87.9 $\pm$ 2.9 & 
			780 $\pm$ 22 & 11600 $\pm$ 2700 \\
			5 & 
			68.4 $\pm$ 0.2 & 54.5 $\pm$ 0.3 & 49.1 $\pm$ 1.2 & 60.7 $\pm$ 2.8 & 
			1000 $\pm$ 92 & 14200 $\pm$ 2800 \\
			6 & 
			79.0 $\pm$ 0.2 & 62.1 $\pm$ 0.3 & 50.7 $\pm$ 1.2 & 62.6 $\pm$ 2.7 & 
			553 $\pm$ 16 & 14900 $\pm$ 2700 \\
			7 & 
			67.2 $\pm$ 0.2 & 50.1 $\pm$ 0.3 & 62.5 $\pm$ 1.1 & 81.4 $\pm$ 2.8 & 
			1190 $\pm$ 256 & 20800 $\pm$ 3100 \\
			8 & 
			93.0 $\pm$ 0.2 & 69.8 $\pm$ 0.3 & 68.9 $\pm$ 1.1 & 56.4 $\pm$ 2.5 & 
			849 $\pm$ 165 & 9900 $\pm$ 2500 \\
			9 & 
			57.2 $\pm$ 0.5 & 51.6 $\pm$ 0.5 & 35.1 $\pm$ 1.2 & 53.3 $\pm$ 2.7 & 
			573 $\pm$ 179 & 16100 $\pm$ 2300 \\
			10 & 
			91.2 $\pm$ 0.2 & 71.3 $\pm$ 0.3 & 46.5 $\pm$ 1.1 & 53.4 $\pm$ 2.9 & 
			590 $\pm$ 199 & 7300 $\pm$ 1800 \\
		\hline
		\end{tabular}
		\label{t_flux2}
		\begin{flushleft}
		\medskip
			\textbf{Note.} - Total fluxes are given in units of $\mu$Jy.\\
		\end{flushleft}
	\end{minipage}
\end{table*}

The sample consists of 12 infrared galaxies at $z = 0.85 - 1.44$. 
The redshifts and data sources of our 12 galaxies are shown in Table \ref{t_sample}. 
They are SST J1604+4304, HR10(ERO J164502+4626.4), and 10 galaxies found 
in the Cosmological Evolution Survey (COSMOS) field.
We have searched archives and the literature for those 
galaxies which meet the following criteria: 
(1) the far-IR emission is detected; (2) the optical spectrum indicates 
a redshift of $\sim$ 1, and gaseous emission lines such as [OII]$\lambda3727$ 
and/or stellar absorption lines such as Ca H/K and HI absorption lines were 
detected; (3) broad-band photometry has been done from the UV to mid-infrared; 
(4) there is no contamination from foreground galaxies; (5) the bolometric 
luminosity is dominated by the undergoing star formation with the criterion of 
$f_{\nu}(8\micron)/f_{\nu}(4.5\micron) < 1.9$ \citep{Pope2008} and
no obvious evidence of emission caused by an AGN such as X-ray emission.
As a result, we have found these 12 galaxies all of which are classified as 
starburst-dominated ULIRGs.

The data for SST J1604+4304 were taken from \citet{Kawara2010} where the 
multi-wavelength data covering the X-ray, UV/optical, infrared, and radio and 
the optical spectrum are given. For HR10, \citet{Stern2006} compiled the 
multi-wavelength data from the UV/optical to radio and the optical spectrum 
(strictly speaking from 2750 to 3850 \AA) is given in \citet{Dey1999}.

The photometric and spectroscopic data for the 10 COSMOS galaxies are available 
in the COSMOS archive at http://irsa.ipac.caltech.edu/data/COSMOS/. Figure \ref{f_photo} 
illustrates postage stamp images of the COSMOS galaxies from $u^*$ to 70\micron.
For full details of the observations of the COSMOS galaxies, 
the reader refers to \citet{Capak2007} for optical to near-infrared imaging performed 
on the {\it Hubble} Space Telescope and ground based telescopes including the Subaru 8.3m telescope, 
the Kitt Peak National Observatory (KPNO) 4m telescope, the Cerro Tololo Inter-American (CTIO) 4m 
telescope, and the Canada-France-Hawaii (CFH) 3.6m telescope, 
to \citet{Sanders2007} for the COSMOS {\it Spitzer} survey (S-COSMOS), \citet{Lefloch2009} for 
the 24\micron\ observations, \citet{Frayer2009} for the 70 and 160 \micron\ observations, 
\citet{Trump2009} for 5600-9200\AA\ spectroscopy on the the Magellan 6.5m telescope, 
and \citet{Lilly2007} for 5500-9700\AA\ spectroscopy on the European Southern Observatory 
(ESO) 8m Very Large Telescope(VLT). 

For COSMOS galaxies, the optical counterparts of far-IR sources were selected as follows.
We first selected objects at $z = 0.8 - 1.2 1$ in the spectroscopy catalogs and searched for the 
counterparts in the optical imaging data catalogs. 
If there is a single object in the optical imaging 
data within an error circle radius of 0.2\arcsec from the spectroscopic position, we identified 
the spectroscopic object with the object in the optical imaging catalogs. 
We then searched for the mid-infrared counterpart in the IRAC (3.6, 4.5, 5.6, 8.0\micron) catalog.
Again, if there is one IRAC object within an error circle radius of 0.5\arcsec from the 
position of the optical, spectroscopic object, we regard the IRAC object as optically 
identified. This process was repeated for the 24\micron\ counterpart within an error radius of 
1.0\arcsec, and 70\micron\ counterpart within 4.0\arcsec. 
We found no 160\micron\ counterparts within 8.0\arcsec. 
As a result, we found 10 starburst-dominated ULIRGS 
at z = 0.8 - 1.2. \citet{Kartaltepe2010}, applying more relaxed criteria than ours, 
found 50 70\micron\ sources having spectroscopic data in the same field and 
the same redshift range. Assuming that 70\micron\ sources with $>$ 10 mJy are ULIRGs 
at $z >$ 0.8, 25 of the 50 sources are ULIRGs with spectroscopic data. 
They suggest that an AGN fraction is about 50\%, thus starburst-dominated ULIRGs 
are about 12, fairly consistent with the number of ULIRGs in our sample.
 
\begin{figure*}
	\begin{minipage}{180mm}
\centering
\includegraphics[scale=0.62,angle=90,clip]{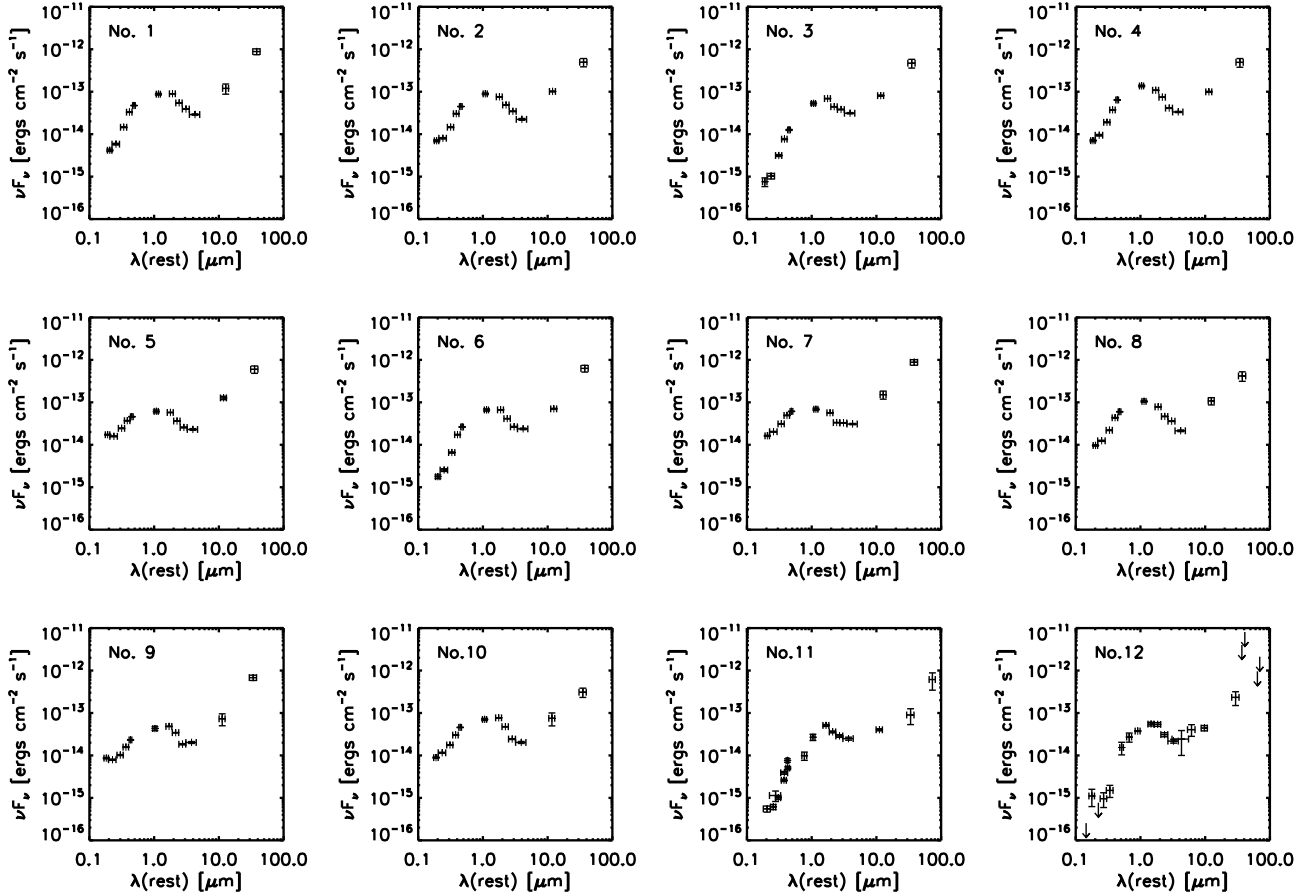}
\caption{0.1 - 100 \micron\ SEDs of the 12 galaxies, 
         illustrating $\nu f_{\nu}$(ergs cm$^{-2}$ s$^{-1}$) as a function of 
         rest wavelength(\micron). 
		}
\label{f_sed}
\end{minipage}
\end{figure*}

Optical data were taken in many photometric bands for COSMOS galaxies. If we use all these data 
in the following analysis, the weight of the spectral energy distribution (SED) in the optical will 
overwhelm the other data. We thus use $u^*$ taken on the CFH telescope, $g^+, r^+, i^+, z^+$ on the 
Subaru telescope, $Ks$ on the CTIO 4m telescope. Aperture photometry with a 3\arcsec diameter is 
given in the catalog. We applied the correction to the total magnitude as instructed by 
\citet{Capak2007}.
The aperture photometry with a 2.9\arcsec diameter given in the S-COSMOS IRAC 4-channel Catalog was 
converted to the total fluxes by applying correction factors of 0.90, 0.90, 0.84, \& 0.73 for 
3.6, 4.5, 5.8, \& 8.0\micron, respectively (see README attached to this catalog). The  
total fluxes of the COSMOS galaxies are summarized in Tables \ref{t_flux1} and \ref{t_flux2} 
in units of $\mu$Jy. The SEDs of the 12 galaxies are plotted as a function of 
rest wavelength in Figure \ref{f_sed}.

\begin{figure*}
	\begin{minipage}{180mm}
		\centering
\includegraphics[scale=0.62,angle=90,clip]{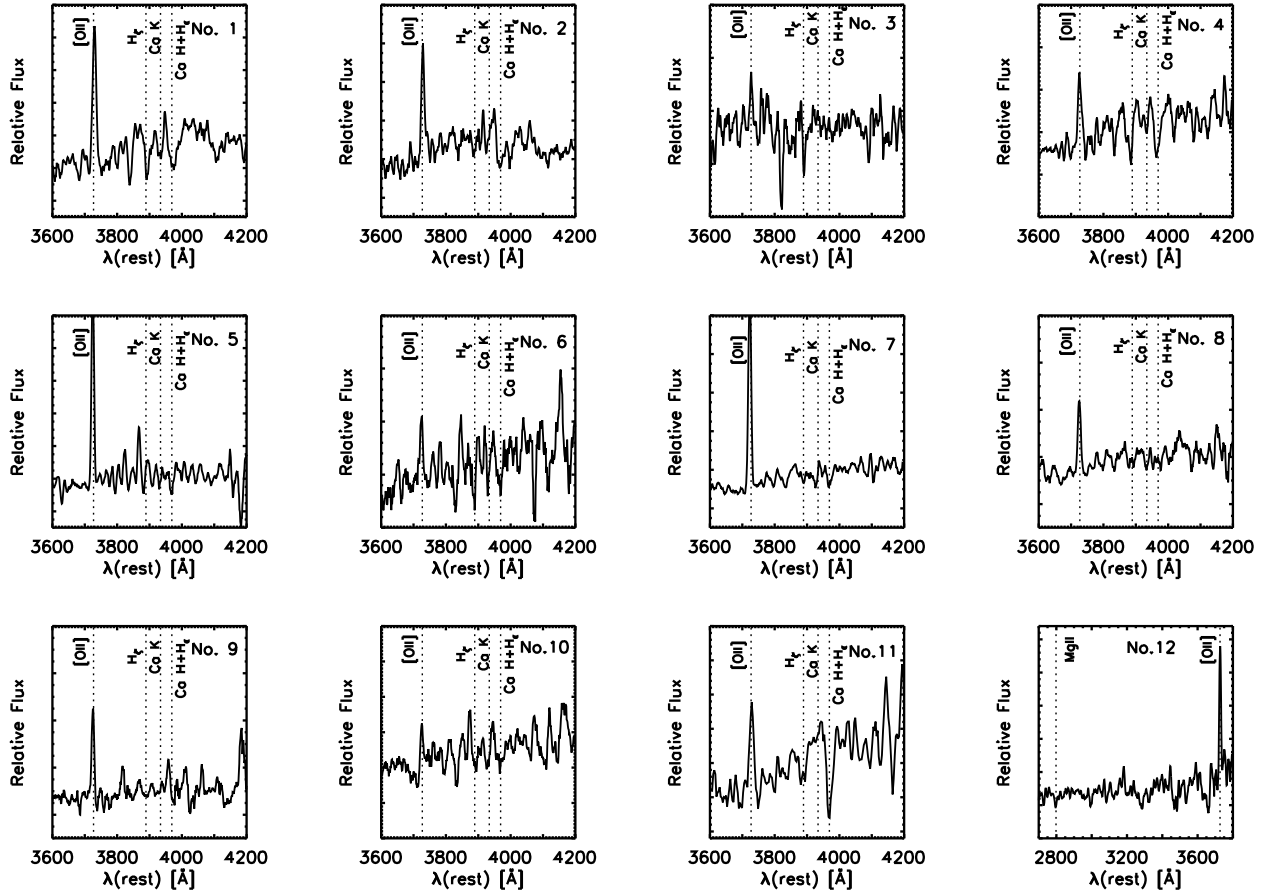}
\caption{Optical spectra of the 12 galaxies. The relative flux densities (per \AA)
are plotted as a function of rest-frame wavelength (\AA). The vertical lines mark the 
positions of MgII at 2798\AA, [OII] at 3727\AA, CaII H \& K at 3968 and 3934\AA\ respectively,
H$_\zeta$ at 3889\AA, and H$_\epsilon$ at 3970 \AA.
		}
\label{f_spec}
\end{minipage}
\end{figure*}

The optical spectra of our 12 galaxies are illustrated in Figure \ref{f_spec} along with 
the vertical lines showing the positions of the [OII]$\lambda$3727 emission line 
from HII regions and 
the stellar absorption lines, namely, MgII$\lambda$2798, CaII H \& K, H$_\zeta$, and H$_\epsilon$. 
[OII]$\lambda$3727 indicates the presence of OB stars, while the absorption lines are useful to 
measure the age of the relatively old stellar populations \citep{Kawara2011}. The equivalent widths 
of the lines are given in Table \ref{t_line} along with the D(CaII) index. This index is defined as 
D(CaII) = 2 $\times$ EW(CaII K)/[EW(H$_{\zeta}$)+EW(CaII H + H$_{\epsilon}$)] \citep{Kawara2010}, 
where EW(CaII K), EW(H$_{\zeta}$), and EW(CaII H + H$_{\epsilon}$) are the equivalent widths of 
CaII K, H$_{\zeta}$, and the blend of CaII H and H$_{\epsilon}$, respectively.

\begin{table*}
	\begin{minipage}{150mm}
	\centering
	\caption{Spectroscopic proterties of the 12 galaxies}
	\begin{tabular}{cccccc}
	\hline
 (1) & (2)         & (3)     & (4)                 & (5)            & (6)                     \\
 No. & EW([OII])   & D(CaII) & EW(H$_\zeta$)       & EW(Ca K)       & EW(Ca H + H$_\epsilon$) \\
     & (\AA)       &         & (\AA)               & (\AA)          & (\AA)                   \\
	\hline
1      & 45.6 $\pm$ 2.4 & 0.28 $\pm$ 0.13 & 10.6 $\pm$ 1.6 & 3.5 $\pm$ 1.5 & 14.2 $\pm$ 2.2 \\
2      & 27.3 $\pm$ 2.7 & 0.11 $\pm$ 0.13 & 4.3 $\pm$ 1.9  & 1.0 $\pm$ 1.1 & 12.8 $\pm$ 1.5 \\
3      & 10.1 $\pm$ 3.5 & 0.27 $\pm$ 0.36 & 12.6 $\pm$ 5.2 & 2.6 $\pm$ 3.4 & 6.8 $\pm$ 3.8 \\
4      & 31.4 $\pm$ 3.3 & 0.15 $\pm$ 0.05 & 6.5 $\pm$ 1.4  & 1.6 $\pm$ 0.6 & 15.4 $\pm$ 1.2 \\
5      & 65.3 $\pm$ 1.8 & 0.18 $\pm$ 0.22 & 6.4 $\pm$ 2.6  & 1.0 $\pm$ 1.2 & 4.7 $\pm$ 1.9 \\
6      & 9.8 $\pm$ 1.8  & 0.51 $\pm$ 0.26 & 8.4 $\pm$ 3.7  & 4.9 $\pm$ 2.2 & 10.8 $\pm$ 2.6 \\
7      & 35.0 $\pm$ 0.6 & 0.13 $\pm$ 0.26 & 1.8 $\pm$ 0.6  & 0.3 $\pm$ 0.6 & 2.9 $\pm$ 0.7 \\
8      & 23.4 $\pm$ 1.3 & 0.18 $\pm$ 0.41 & 3.6 $\pm$ 1.0  & 0.6 $\pm$ 1.2 & 2.6 $\pm$ 2.3 \\
9      & 63.2 $\pm$ 6.0 & 0.02 $\pm$ 0.14 & 6.1 $\pm$ 1.9  & 0.1 $\pm$ 1.0 & 9.0 $\pm$ 1.9 \\
10     & 9.1 $\pm$ 2.1  & 0.37 $\pm$ 0.26 & 5.8 $\pm$ 2.0  & 2.6 $\pm$ 1.8 & 8.4 $\pm$ 2.4 \\
11$^a$ & 31.1 $\pm$ 4.3 & 0.03 $\pm$ 0.11 & 7.6 $\pm$ 2.1  & 0.4 $\pm$ 1.4 & 16.7 $\pm$ 2.2 \\
12$^b$ & 47 $\pm$ 5 & \multicolumn{4}{c}{EW(MgII) = 7.6 $\pm$ 3.1} \\
	\hline
	\end{tabular}
	\label{t_line}
	\begin{flushleft}
	\medskip
	\textbf{Note.} - Col 2 for the eqivalent width of the [OII] line at 3727\AA. 
        Col 3 for the CaII index defined as D(CaII) = 
        2 $\times$ EW(CaII K)/[EW(H$_{\zeta}$)+EW(CaII H + H$_{\epsilon}$)] \citep{Kawara2010}.
        Cols 4-5 for the equivalent width for the H$_{\zeta}$ line at 3889\AA\ and 
        the CaII K line at 3934\AA, respectively. 
        Col 6 for the equivalent width of the blends of the CaII H \& H$_{\epsilon}$ lines at 
        3968 and 3970\AA, respectively.\\
        $^a$ The equivalent widths were taken from \citet{Kawara2010}. \\
        $^b$ EW([OII]) was taken from \citet{Dey1999},and EW(MgII$\lambda2798$) was measured 
        using the spectrum given by \citet{Dey1999}. 
	\end{flushleft}
	\end{minipage}
\end{table*}

Table \ref{t_lum} gives the 8-1,000\micron\ luminosity $L_{IR}$ and the mass of dust $M_d$ of the 
12 sample galaxies. We obtained $L_{IR}$ by fitting the \citet{Dale2002} ULIRG template to 
the 70 \micron\ data except for 
SST J1604+4304 for which the 70 and 160 \micron\ data were used for fitting. 
Deriving far-IR colors $f_\nu(160\micron)/f_\nu(70\micron)$ of 70\micron\ sources that are 
not detected at 160\micron\ from the stacking of the 160\micron\ data points, 
\citet{Kartaltepe2010} found that those not 
detected at 160\micron\ have hotter far-IR colors than those detected at 160 \micron, and that 
the far-IR colors of 70\micron\ sources not detected at 160 \micron\ scatter around 
the \citet{Dale2002} ULIRG template 
at $z \sim 1$. It is thus appropriate to utilize this template to derive 
$L_{IR}$ for those not detected at 160 \micron. The typical uncertainty for $L_{IR}$ using 
the 70\micron\ data is estimated to be $\sim 0.2$ dex \citep{Kartaltepe2010}.
SST J1604+4304 was re-analyzed in the same way as the other galaxies. 
While the Arp 220 template was used to estimate $L_{IR}$ and the dust temperature in our previous 
analysis \citep{Kawara2011}, we apply the \citet{Dale2002} ULIRG template in the present analysis.
In addition, we include a new term of the dust mass in the minimum $\chi^2$ method. The 
new results are quite consistent with our previous results.
 
The dust mass was estimated using the formula of 
$M_d/f_{\nu} = D_L^2/[(1+z)\kappa_{\nu_e}B_{\nu_e}(T_d)]$,where $f_{\nu}$ is the observed 
flux density at frequency $\nu$, $\nu_e = (1+z)\nu$, $D_L$ is the luminosity 
distance to the emitter, $B_{\nu_e}(T_d)$ is the Planck function for $T_d$ the temperature of dust 
at $\nu_e$, $\kappa_{\nu_e}$ is the absorption cross-section per unit dust mass at $\nu_e$. 
As can be seen in this formula, $M_d$ depends on the product 
$\kappa_{\nu} B_{\nu}(T_d)$ $\propto$ $\nu^{\beta} B_{\nu}(T_d)$ 
if $\kappa_{\nu}$ $\propto$ $\nu^{\beta}$.

It is known that there is an anti-correlation between $\beta$ and $T_{d}$ (\citealt{Blain2003}; 
\citealt{Bracco2011}). This degeneracy may be due to the intrinsic 
properties of dust grains or application of a single-temperature model to the SEDs which 
average various temperatures along the line of sight \citep{Bracco2011}. Thus, when fitting a 
single-temperature model to the SED, 
the resultant $T_{d}$ depends on $\beta$ or the properties of the dust extinction curves; 
the extinction curves we study here, are the MW, pseudo MW, Calzetti, SN curves, 
as fully described in section 3.1.2.

The dust temperatures cannot be directly derived from our data, 
because of the lack of data longward of 160 \micron. 
We thus estimated $T_d$ by fitting to the \citet{Dale2002} ULIRG template, 
resulting in $T_d = 40.5 \pm 9.3$ for the SN dust with $\beta$ $\simeq 1.6$
and $T_d = 37.0 \pm 8.9$ for the MW dust with  $\beta$ $\simeq 2 $; 
$T_d$ derived with the SN dust is 10\% higher than that with the MW dust.  This temperature 
difference is expected from the degeneracy between $\beta$ and $T_{d}$.
Table \ref{t_lum} gives $M_d$ for the 8-40$M_\odot$ SN dust.
$M_d$ in Table \ref{t_lum} should be multiplied by 0.81 for the 8-30$M_\odot$ SN dust 
and 0.78 for the MW dust. It is noted that 
$M_d$ is derived by using $\kappa_{\nu_e}$ at 80\micron\ which 
is 44cm$^2$ g$^{-1}$ for MW dust,24cm$^2$ g$^{-1}$ for 8-30$M_\odot$ SN dust, 
and 20cm$^2$ g$^{-1}$ for 8-40$M_\odot$ SN dust (see Figure 7 in \citealt{Kawara2011}).

\begin{table}
\centering
\caption{Infrared luminosity, dust mass, and dust shell radius 
of the 12 galaxies}
	\begin{tabular}{cccc}
	\hline
	No. & 
	$L_{IR}$$^a$ ($10^{12}L_\odot$) & $M_d$$^b$ ($10^{8}M_\odot$) & 
	$r$$^c$ (arcsec) \\
	\hline
		1 & 2.60 $\pm$ 1.14 & 1.66 $\pm$ 1.06 & 0.63 \\
		2 & 2.87 $\pm$ 1.25 & 1.55 $\pm$ 1.02 & 0.74 \\
		3 & 2.66 $\pm$ 1.18 & 1.73 $\pm$ 1.15 & 0.47 \\
		4 & 2.65 $\pm$ 1.20 & 2.15 $\pm$ 1.43 & 0.52 \\
		5 & 3.73 $\pm$ 1.63 & 2.00 $\pm$ 1.30 & 0.80 \\
		6 & 1.63 $\pm$ 0.74 & 1.42 $\pm$ 0.91 & 0.61 \\
		7 & 2.78 $\pm$ 1.21 & 1.65 $\pm$ 1.05 & 0.85 \\
		8 & 1.82 $\pm$ 0.81 & 0.94 $\pm$ 0.63 & 0.61 \\
		9 & 2.93 $\pm$ 1.39 & 3.30 $\pm$ 2.11 & 0.62 \\
		10 & 2.12 $\pm$ 0.95 & 1.16 $\pm$ 0.78 & 0.60 \\
		11 & 1.04 $\pm$ 0.50 & 1.94 $\pm$ 0.91 & 0.48 \\
		12 & 7.05 $\pm$ 3.25 & 3.72 $\pm$ 2.69 & 0.47 \\
	\hline
	\end{tabular}
	\label{t_lum}
	\begin{flushleft}
	\medskip
$^a$ $L_{IR}$ is the 8-1,000\micron\ bolometric infrared luminosity. \\
$^b$ The 8-40$M_\odot$ SN dust is assumed.
For the MW dust, a factor of 0.78 should be multiplied. \\
$^c$ $r$ is the radius of dust shell. It can be apprpximated as 
$r\sim\sqrt{r_x r_y}$ which were obtained from HST I$_{814}$ images.
	\end{flushleft}
\end{table}

\section{Analysis}

We take a similar approach as performed in our previous study. 
Synthetic spectra of stellar populations reddened by various dust models are applied 
to reproduce the data of the 12 sample galaxies. For the full details of this approach, 
the reader refers to \citet{Kawara2011} and references therein.
 
\subsection{Dust}

Our dust models are a combination of an optical properties of dust 
and a geometry of the dust distribution.

\subsubsection{Geometries of dust distribution}

The attenuation produced by dust in front of an extended source depends not only on the 
optical properties of dust grains, but also on the geometry of 
spatial distribution of dust, especially when starlight is attenuated by a thick, 
clumpy dust layer or dust is internal of the extended source. Analytic solutions to 
determine the dust parameters for various geometries are extensively studied by \citet{Natta1984} 
and \cite{Code1973}. \citet{Code1973} applied the two-stream approximation to the radiative 
transfer for a circumstellar dust shell and this approximation was developed for a dust slab 
containing emitters and absorbers in Appendix by \citet{Kawara2011}. It is noted that 
the slab approach is a good approximation to a spherical geometry.
The reader may also refer to an excellent review and discussion in 
\citet{Calzetti1994}.
  
We consider the following geometries of dust distribution:
(1) {\it Internal dust geometry} -- absorbers(dust grains) and emitters(stars) are uniformly 
mixed and distributed; 
(2) {\it Foreground dust screen geometry} -- absorbers are physically separated from 
and located in a screen in front of emitters, where we assume that there are $N$ clumps in average 
along the line of sight, thus the dust distribution is clumpy (small $N$) or 
uniform (very large $N$);
(3){\it Composite geometry} -- dust grains distribute both within the system of 
stars as internal dust and in the screen in front of the system. 

The effects of dust also depend on the physical distance between the dust and the stars 
\citep{Calzetti1994}. 
If the dust is located physically distant from the stars, 
the effect of dust is to remove photons from the line of sight 
through absorption and scattering. Hereafter, we call this distant configuration {\it no scattering}.
On the other hand, if the dust is located close to the stars, 
the effect is not only to remove photons from the line of sight but also to convey photons 
back into the line of sight by scattering. We examine this close configuration in two extreme cases, 
namely, {\it forward-only scattering} and {\it isotropic scattering}.
  In the internal dust geometry where the dust is located close to the stars, 
we account for the effect of the scattering through the relations 
$\tau_{eff} = (1-\gamma)\tau_{ext}$ for  forward-only scattering and 
$\tau_{eff} = (1-\gamma)^{0.75}\tau_{ext}$ for isotropic scattering\citep{Kawara2011}, 
where $\gamma$ is albedo, $\tau_{eff}$ the effective absorption optical depth, 
$\tau_{ext}$ the sum of the optical depth of absorption and scattering.
In the foreground screen geometry, if the dust screen is physically distant from the stars, 
no scattering applies. On the other hand, if the dust screen is physically 
close to the stars, the scattering can put photons into the line of sight.
We thus consider three cases for the foreground geometry: $\tau_{eff} = \tau_{ext}$ 
for no scattering, $\tau_{eff} = (1-\gamma)\tau_{ext}$ for  forward-only scattering, and 
$\tau_{eff} = (1-\gamma)^{0.5}\tau_{ext}$ for  isotropic scattering. 
The relation between the observed flux $f_{\nu}$ and the unobscured flux $f^0_{\nu}$ is:
$f_{\nu}/f^0_{\nu} =  (1-e^{-\tau_{eff}})/\tau_{eff}$ for an internal dust geometry and  
$f_{\nu}/f^0_{\nu} =  e^{-\tau_{ap}}$ with $\tau_{ap} = N(1-e^{-\tau_{eff}/N})$ for a dust screen 
geometry with the number of $N$ clumps along the line of sight (\citealt{Natta1984}). 
It is noted that there is a definite difference between 
attenuation and extinction curves; we regard attenuation curves as  $f_{\nu}/f^0_{\nu}$ and
extinction curves as $\tau_{ext}$. 

\subsubsection{Extinction curves}

We examine two extinction curves synthesized in ejecta of SNe II, the empirical extinction curve 
derived by  \citet{Calzetti2001}, the MW extinction curve given by \citet{Draine2003}, 
and the pseudo MW extinction curve. 

The MW extinction curve exhibits a pronounced feature at 2175 \AA. 
This feature is known to be absent 
in the typical extinction curves of the Small Magellanic Clouds (SMC) and some galaxies. 
To check the effect of the feature on the extinction law in our galaxies, we create 
the pseudo MW extinction curve which is made by removing the 2175 \AA\ feature, namely, 
the extinction bump and albedo dip, from the MW extinction curve.

The Calzetti extinction curve is based on the attenuation law that 
was derived in a range from 0.12 to 2.2 \micron\ by analyzing local starbursts and blue compact 
dwarf galaxies \citep{Calzetti1994}. \citet{Calzetti2001} expresses this law in a functional form 
combined with a geometry of a uniform dust screen with no scattering 
(i.e., $f_{\nu} = e^{-\tau_{\nu}}f^0_{\nu}$). This cannot be used for longer wavelengths, 
because the extinction curve ($\tau_{\nu}$) sharply drops beyond 2.2 \micron. The Calzetti
extinction curve is almost identical to the MW curve from 0.6 to 1.6 \micron.  
We thus extend the Calzetti curve to longer wavelengths
by adopting the MW extinction curve for wavelengths $ \ge$ 1.6 \micron.  
Because the Calzetti curve has already been processed by the foreground screen, 
we consider this geometry only for the Calzetti curve.

Combining the optical constants in the literature with the grain species and size distribution of 
SN-condensed dust derived by \citet{Nozawa2003}, and using the Mie theory, 
\citet{Hirashita2005} calculated absorption and scattering cross 
sections of homogeneous spherical grains with various sizes condensed in mixed and unmixed ejecta.
Comparing their models to the extinction curve of SDSS J1048+637 (a BALQSO at $z \sim 6.2$), 
\citet{Hirashita2005} suggest the mass range of SN progenitors to be $\lesssim$ 30 M$_{\sun}$, 
whereas the upper mass of SN progenitors 
is usually assumed to be 40 M$_{\sun}$. We thus adopt the two extinction curves of 
SNe II for the \citet{Salpeter1955} Initial Mass Function (IMF); 
one has a mass range of progenitors from 8 to 30 $M_{\sun}$, and the other 
from 8 to 40 $M_{\sun}$. These were obtained by 
interpolating or extrapolating the extinction curves that \citet{Hirashita2005} 
calculated for unmixed SNe II with progenitor masses of 13, 20, 25, \& 30 $M_{\sun}$. 
We adopt these in the present study and call the original SN extinction curves in a sense 
that the subsequent processing of dust destruction are not considered.
Recent studies indicate that the original SN extinction curves become flatter through 
the reverse shock destruction (\citealt{Bianchi2007}; \citealt{Hirashita2008}), 
while \citet{Hirashita2010} shows the importance of the shattering in a warm ionized medium 
(WIM) in starburst environments and this effect makes the extinction curves steeper. 
They suggest that the final extinction curves, after the reverse shock destruction and 
the shattering, have a slope similar to that of the original curve. The degree of the 
reverse shock destruction and shattering critically depends on the assumed parameters, 
especially gas density of the ISM. Considering these uncertainties and the similarity between 
the original and final extinction curves, we decided to adopt the original SN extinction curves 
in the present study.

\subsection{Star formation history}

\begin{table}
	\centering
	\caption{Parameters of synthetic template  spectra.}
	\setlength{\tabcolsep}{0.3pt}
	\begin{tabular}{cl}
	\hline
Parameter & \hspace{10mm} Range \\
	\hline
Z            & 0.005 - 2.5$Z_\odot$ at 0.005,0.02,0.2,0.4,1.0,2.5$Z_\odot$ \\
t                & 1 - 6,300 Myr    in a step of 0.2 dex  \\
$t_{sfr}$        & 10 - 10,000 Myr in a step of 0.3 dex \\
$A_{0.3\micron}$ & 0 - 15 mag in a step 0.1 mag  \\
                 & 15 - 100 mag in a step of 1 mag \\
	\hline
	\end{tabular}
	\label{t_para}
	\begin{flushleft}
	\medskip
\textbf{Note.} - Z, t, $t_{sfr}$, and $A_{0.3\micron}$ are 
the metallicity, age after the onset of the star formation in Myr, 
e-folding timescale of the star formation in Myr, 
and extinction at 0.3\micron\ in magnitude, respectively. 		
	\end{flushleft}
\end{table}

The evolutionary synthesis codes by \citet{Bruzual2003} are 
used to generate synthetic spectra of evolutionary stellar population models.
We adopt the \citet{Salpeter1955} IMF with the mass cutoffs of 0.1-100M$_\odot$.
The star formation histories are exponentially declining models 
with the star formation rate (SFR) $\propto t^{-1}_{sfr}\exp(-t/t_{sfr})$, 
where $t_{sfr}$ is the e-folding timescale and t is the age.
The synthetic spectra are generated in the parameter range as shown in Table \ref{t_para}.

The equivalent widths of [OII]$\lambda$3727 and MgII$\lambda$2798, 
and D(CaII) were calculated from the synthetic spectra.
[OII] is a nebular line and unpredictable directly from the synthetic spectra.
We first obtain the H$_\alpha$ luminosity by counting ionizing photons in the synthetic spectra.
The metallicity scaling of [OII]/H$_\alpha$ for the ionized parameter $q = 3\times 10^7$cm s$^{-1}$ 
was given by \citet{Kewley2004}.
We thus adopt their theoretical curves with an additional condition, 
which is applied to avoid a negative or unrealistic value, of [OII]/H$_\alpha > 0.1$.
\citet{Kewley2004} estimated the uncertainty in the [OII] intensity to be $\sim$30\%.

\subsection{Minimum $\chi^2$ method}
 We search for the best set of parameters by using a minimum $\chi^2$ method as defined below:
\begin{equation}
\chi^2 = \chi^2_{SED} + W(\chi^2_{\text{[OII]}} 
       + \chi^2_{\text{EW(abs)}} + \chi^2_{L_{IR}} + \chi^2_{M_d}). \label{e_chi2}
\end{equation}

The weighting factor $W$ is introduced to balance the weight between the SED data\footnote{
The number of photometric data points was chosen so that the SED fitting did not overweight 
the other data. However, we found at a late stage of the analysis that the weight of the SED 
still overwhelmed the other data for many galaxies. $W$ was thus introduced to ensure that 
the $\chi^2$ of the SED has the similar magnitude as that of the other data. As a result,
we obtained $W$ = 1 for galaxies 2, 4, 6, 12, and $W$ = 3 for the others}. 

Each term is defined as:
\begin{align}
\chi^2_{SED} &= \sum^{N_{filters}}(F_{obs}-bF_{temp})^2/\sigma^2_{SED}, \label{e_sed} \\
\chi^2_{\text{[OII]}} &= (EW([\text{OII}])_{obs}-EW([\text{OII}])_{temp})^2/\sigma^2_{\text{[OII]}}, 
    \label{e_oii} \\
\chi^2_{\text{EW(abs)}} &= (EW(\text{abs})_{obs}-EW(\text{abs})_{temp})^2/\sigma^2_{EW(\text{abs})} 
    \label{e_abs} \\
\chi^2_{L_{IR}} &= (L_{IR}-L_{IR\ temp})^2/\sigma^2_{L_{IR}}, \label{e_lir} \\
\chi^2_{M_d} &= (M_{d}-M_{d\ temp})^2/\sigma^2_{M_d}. \label{e_mdust}
\end{align}

Equation \ref{e_sed} is related to photometric data of the continuum from stellar atmosphere.
$b$ is a normalization constant between the template flux density $F_{temp}$ and 
the observed $F_{obs}$, and the number of photometric filters $N_{filters}$ from UV to 5.8 \micron.
8.0\micron\ data are excluded from the SED data, 
because dust emission may be significant in the 8.0\micron\ band.
In order to avoid too small and non-realistic photometric error,
5\% of the flux is added in the quadratic form to the photometric error $\sigma_{ph}$, 
i.e., $\sigma^2_{SED} = \sigma_{ph}+(0.05F_{obs})^2$.

Equation \ref{e_oii} is related to the [OII] line from HII regions ionized by young OB stars.
$EW([OII])$ predicted using \citet{Kewley2004} theoretical curve contains $\sim$30\% 
uncertainty, thus 
$\sigma^2_{\text{[OII]}} = \sigma^2_{EW(\text{[OII]})obs}+(0.3EW(\text{[OII])}_{temp})^2$.
where $\sigma^2_{EW(\text{[OII]})obs}$ is the observed error.
Equation \ref{e_abs} measures the absorption lines in stellar atmosphere.
$EW(abs)$ is $EW(MgII)$ for HR10 and $D(CaII)$ for the others.$\sigma^2_{EW(\text{abs})}$ 
is the observed error. 

Equation \ref{e_lir} compares $L_{IR}$ given in Table \ref{t_lum} with the prediction from 
the model, i.e., $L_{IR\ temp} \propto \int \limits(f_{\nu}^0 -f_{\nu})d\nu$, 
where $f_{\nu}^0$ and $f_{\nu}$ are intrinsic and reddended synthetic spectra.
Equation \ref{e_mdust} compares $M_d$ derived from fir-IR data
with $M_{d\ temp}$ the dust mass estimated from the dust extinction (i.e., optical depth) and the 
assumed dust geometry. $M_{d\ temp}$ was derived as follows. 
We first assume a composite geometry consisting of screen dust and internal dust.
Suppose a spherical shell with a radius of $r$ and a thickness of $\delta r$, 
where the internal and screen dust fills inside and in the shell, respectively. Then,
assuming $\delta r/r \ll 1$,
the mass of the internal dust is $M_{in} = 4\pi(\tau_{in}/\kappa_{ext})r^2/3$, 
while that of the screen dust is $M_{sc} = 4\pi(\tau_{sc}/\kappa_{ext})r^2$. 
$\tau_{in}$ and $\tau_{sc}$ are $\tau_{ext}$ for internal 
dust and screen dust, respectively, and they are related to $r$ and $\delta r$ 
as $\tau_{in} = \rho_{in}\kappa_{ext} r$ and $\tau_{sc} = \rho_{sc}\kappa_{ext} \delta r$,where  
$\rho_{in}$ and $\rho_{sc}$ are the mass density of the internal and screen dust, respectively.
The dust mass is derived directly from $\tau_{in}$, $\tau_{sc}$, and $r$ with no needs to 
assume $\delta r$, $\rho_{in}$ and $\rho_{sc}$.
$r$ can be approximated as $r\sim\sqrt{r_x r_y}$ which were obtained from $HST$ $I_{814}$ images,
where $r_x$ and $r_y$ are the semi-major and the semi-minor axes, respectively. $r$ is given 
in Table \ref{t_lum}.

\section{results}

\begin{figure*}
\includegraphics[scale=0.62,angle=90,clip]{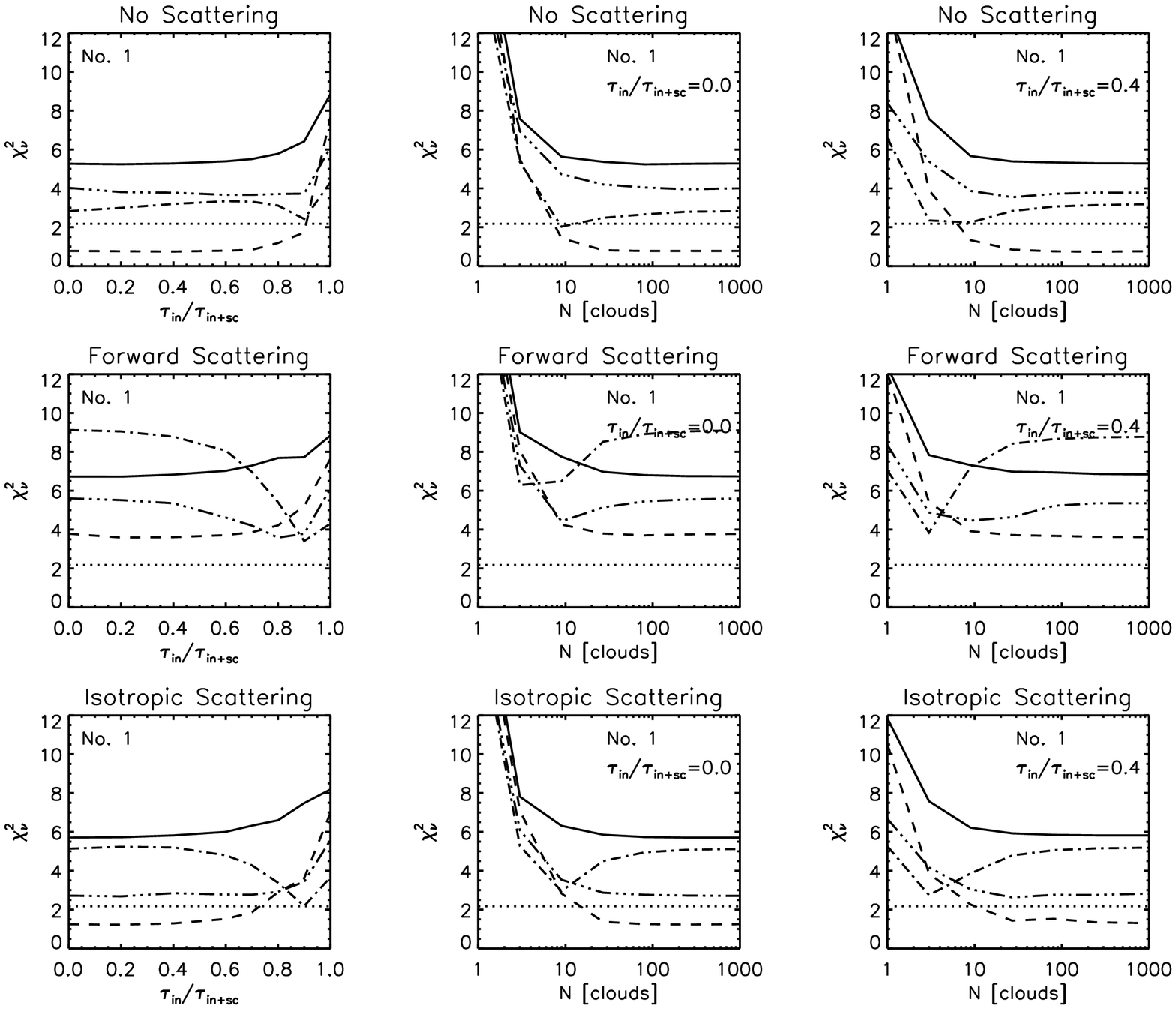}
	\includegraphics[scale=0.6,angle=90,clip]{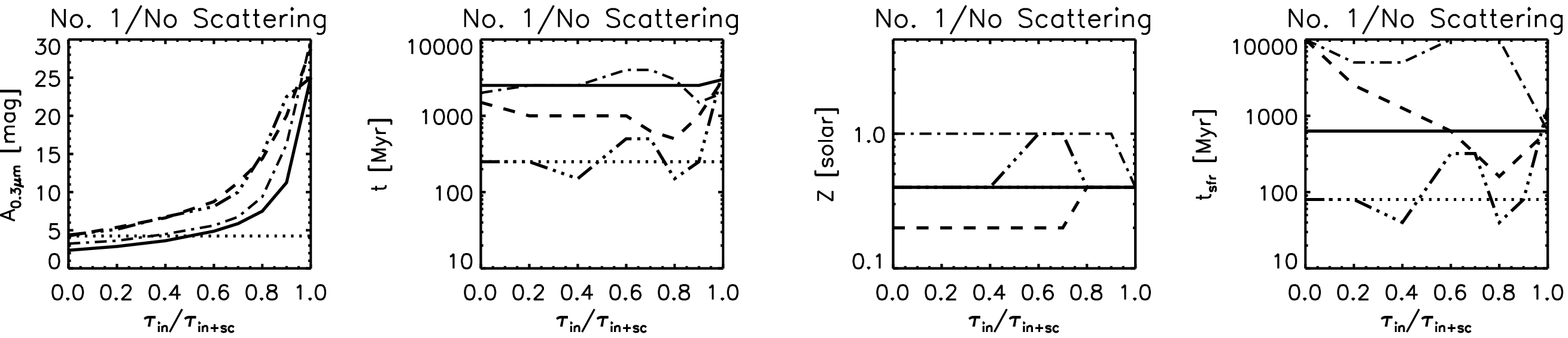} \\
	\includegraphics[scale=0.5,angle=90,clip]{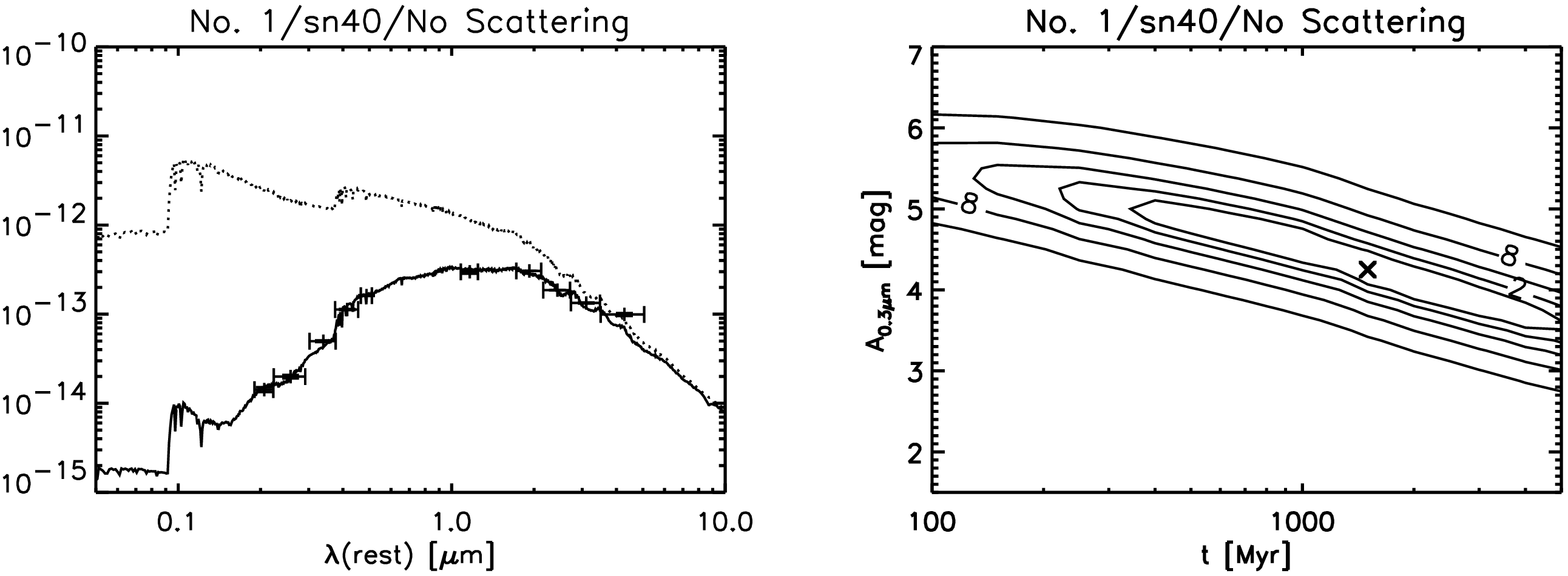}
\caption{Results of the minimun $\chi^2$ analysis. The source number and the scattering property are 
indicated in each panel. The {\it left} panels in {\it first 3 row} show $\chi^2_{\nu}$ for a uniform screen
as a function of $\tau_{in}$/($\tau_{in}+\tau_{sc})$.  The {\it middle} and {\it right} panels in {\it first 3 row} represent 
a screen of $N$ clumps for $\tau_{in}$/($\tau_{in}+\tau_{sc})=0$ and $\tau_{in}$/($\tau_{in}+\tau_{sc})=0.4$, respectively.
In case of no scattering, the scattering properties are no scattering for 
screen dust and forward scattering for internal dust.
{\it Solid} lines represent 8 - 30 M$_{\sun}$ SN dust, 
{\it dash} 8 - 40 M$_{\sun}$ SN dust, {\it dash dot} MW dust, and {\it dash dot dot dot} pseudo 
MW dust, {\it dot} the Calzetti dust. The {\it 4th row} panels show  $\tau_{in} + \tau_{sc}$ 
at 0.3\micron\ in magnitude $A_{0.3\micron}$, 
the metallicity $Z$, the age $t$(yr), and the timescale $t_{sfr}$ as a function of 
$\tau_{in}/(\tau_{in} + \tau_{sc})$.
The {\it bottom row} panels show the best-fit SED along with the unreddended SED and the contour map 
of $\chi_{\nu}^2 - \chi_{\nu}^2(min)$. The full version of the paper is for online publication only, where the full version of 
this figure is available.
}
	\label{f_result}
\end{figure*}

\begin{table*}
	\begin{minipage}{170mm}
	\centering
	\caption{Comparison of dust models}
	\setlength{\tabcolsep}{5pt}
	\begin{tabular}{cclllcccccccc}
	\hline
No. & & \multicolumn{5}{c}{Reduced $\chi$ square ($\chi^2_{\nu}$$^a$)}  & & \multicolumn{5}{c}{Scattering property} \\
    & & \multicolumn{5}{c}{$\tau_{in}/\tau_{in+sc}$ ($N^b$)}  & & \multicolumn{5}{c}{$A_{0.3\micron}$} \\
	        \cline{3-7}  \cline{9-13} 
    & & MW & pMW & Cal & SN30 & SN40 & & MW & pMW & Cal & SN30 & SN40 \\ 
	\hline 
        \hline
1   & & 2.18 & 2.72 & 2.18 & 5.27 & 0.79* & & Isotrop & Isotrop & -- & No      & No \\
    & & 0.9 (--) & 0.0 ($\infty$) & -- & 0.0 ($\infty$) & 0.0 ($\infty$) & & 17.5 & 7.1 & 4.3 & 2.4 & 4.3 \\ 
		  & &     &     &    &       &     &      &      &     &   &     \\	
2   & & 3.35 & 2.87 & 2.04 & 3.65 & 1.61* & & Isotrop & No      & -- & Isotrop & No \\
    & & 0.9 (--) & 0.6 (--) & -- & 0.0 ($\infty$) & 0.0 ($\infty$) & & 14.4 & 7.3 & 3.9 & 4.1 & 3.4 \\ 
		  & &     &     &    &       &     &      &      &     &   &     \\	
3   & & 2.09 & 1.98 & 2.42 & 2.51 & 1.62* & & Isotrop & Isotrop      & -- & No & No \\
    & & 0.8 (--) & 0.0 ($\infty$) & -- & 0.9 (--) & 0.0 ($\infty$) & & 23.8 & 8.8 & 5.9 & 18.8 & 6.0 \\ 
		  & &     &     &    &       &     &      &      &     &   &     \\	
4   & & 3.93 & 0.86* & 3.02 & 8.72 & 2.33 & & Forward & Isotrop      & -- & No & No \\
    & & 0.9 (--) & 0.0 ($\infty$) & -- & 0.0 ($\infty$) & 0.0 ($\infty$) & & 25.0 & 5.9 & 2.8 & 2.5 & 3.5 \\ 
		  & &     &     &    &       &     &      &      &     &   &     \\	
5   & & 1.52 & 2.19 & 2.07 & 2.45 & 0.80* & & No & No      & -- & No & No \\
    & & 0.9 (--) & 0.7 (--) & -- & 0.0 ($\infty$) & 0.0 (10) & & 13.1 & 7.5 & 3.3 & 2.9 & 4.8 \\ 
		  & &     &     &    &       &     &      &      &     &   &     \\	
6   & & 1.88 & 1.72 & 1.61 & 1.51 & 1.37* & & No & Isotrop      & -- & No & No \\
    & & 0.8 (--) & 0.0 ($\infty$) & -- & 0.0 ($\infty$) & 0.0 ($\infty$) & & 11.9 & 6.8 & 4.3 & 2.8 & 4.9 \\ 
		  & &     &     &    &       &     &      &      &     &   &     \\	
7   & & 4.30 & 4.59 & 5.00 & 5.61 & 2.35* & & No & Isotrop      & -- & No & No \\
    & & 0.0 (10) & 0.0 ($\infty$) & -- & 0.0 ($\infty$) & 0.0 ($\infty$) & & 3.3 & 4.4 & 3.0 & 2.6 & 3.6 \\ 
		  & &     &     &    &       &     &      &      &     &   &     \\	
8   & & 1.35 & 1.99 & 1.51 & 3.00 & 0.98* & & No & Isotrop      & -- & Isotrop & No \\
    & & 0.0 (10) & 0.0 ($\infty$) & -- & 0.0 ($\infty$) & 0.0 ($\infty$) & & 4.0 & 5.4 & 3.6 & 2.9 & 3.6 \\ 
		  & &     &     &    &       &     &      &      &     &   &     \\	
9   & & 2.15 & 1.02* & 1.51 & 2.10 & 1.88 & & No & Isotrop      & -- & No & No \\
    & & 1.0 (--) & 0.9 (--) & -- & 0.0 ($\infty$) & 0.0 (10) & & 32.5 & 21.3 & 3.5 & 3.6 & 4.9 \\ 
		  & &     &     &    &       &     &      &      &     &   &     \\	
10  & & 2.88 & 1.23 & 0.75* & 2.37 & 2.08 & & Isotrop & Isotrop   & -- & Foward & No \\
    & & 0.8 (--) & 0.0 ($\infty$) & -- & 0.0 ($\infty$) & 0.0 ($\infty$) & & 9.4 & 4.5 & 3.6 & 4.0 & 3.6 \\ 
		  & &     &     &    &       &     &      &      &     &   &     \\	
11  & & 5.09 & 5.55 & 5.05 & 3.56* & 3.75 & & No & Isotrop     & -- & No & Foward  \\
    & & 0.0 (10) & 0.9 (--) & -- & 0.0 ($\infty$) & 0.0 ($\infty$) & & 8.8 & 42.5 & 5.8 & 5.4 & 17.5 \\ 
		  & &     &     &    &       &     &      &      &     &   &     \\	
12  & & 0.84* & 0.84* & 1.36 & 1.71 & 1.41 & & Isotrop  & Isotrop     & -- & No & No  \\
    & & 0.4 (3) & 0.4 (3) & -- & 0.0 ($\infty$) & 0.0 ($\infty$) & & 20.0 & 20.0 & 5.0 & 4.0 & 4.4 \\	
	\hline
	\end{tabular}
	\label{t_chi2}
	\begin{flushleft}
	\medskip
\textbf{Note.} - The smallest $\chi^2_{\nu}$ is given per galaxy per dust model. "*" marks $\chi^2_{\nu}$ of the 
best-fit model for each galaxy.
'MW', 'pMW', 'Cal', 'SN30', and 'SN40' denote a dust model of the MW, pseudo MW, Calzetti, 8-30$M_\odot$ SN, 
and 8-40$M_\odot$ SN, respectively. 'Isotrop', 'Forward', and 'No' denote isotropic, forward, and no scattering,
respectively.\\ 
$^a$ $\chi^2_\nu=\chi^2_\nu/\nu$, where $\nu$ is degree of freedom.\\
$^b$ The number of clumps $N$ is given in parentheses for a geometry dominated by foreground screen dust.\\
	\end{flushleft}
	\end{minipage}
\end{table*}

The results of the minimum $\chi^2$ analysis are illustrated in Figure \ref{f_result}.
The left panels of the first 3 rows plot $\chi^2_{\nu}$ (=$\chi^2/{\nu}$, 
where ${\nu}$ is degree of freedom)\footnote{
Degree of freedom is defined as $\nu = N_{filter} + 4W -5$ where $N_{filter}$ is the number 
of photometric data used for $\chi^2_{SED}$ and $W$ the weighting factor. The number of fitted
parameters is 5 (i.e., $b$, $t$, $t_{sfr}$, $A_{0.5\micron}$, $Z$);
$N_{filter}$ = 9 for the COSMOS galaxies, 13 for SST J1604, and 10 for HR 10.} 
as a function of $\tau_{in}/(\tau_{in} + \tau_{sc})$ 
for a composite geometry consisting of uniform screen dust and internal dust; 
$\tau_{in}/(\tau_{in} + \tau_{sc})$ = 0 for the foreground screen geometry, 
while $\tau_{in}/(\tau_{in} + \tau_{sc})$ = 1 for the internal dust geometry. 
The scattering properties are assumed to be no scattering, isotropic scattering 
($g$ = 0) , or forward-only scattering ($g$ = 1). In case of no scattering, 
the scattering properties are no scattering for 
screen dust and forward scattering for internal dust because no scattering 
cannot be assumed for internal dust.
The middle and right panels of the first 3 rows show $\chi^2_{\nu}$ as a function of $N$ the number 
of clumps along the line of sight in the foreground screen geometry and the composite geometry with 
$\tau_{in}/(\tau_{in} + \tau_{sc})$ = 0.4. The several linestyles are used to specify the extinction 
curve: {\it Solid} lines to the 8-30 $M_{\sun}$ SN curve, {\it dash} to the 8-40 $M_{\sun}$ SN, 
{\it dot} to the Calzetti, {\it dash dot} to the MW, and {\it dash dot dot dot} to the pseudo MW.

The four panels in the forth row plot $A_{0.3\micron}$ the extinction ($\tau_{in} + \tau_{sc}$) 
in magnitude 
at 0.3\micron, $t$ the age (i.e., the time after the onset of the star formation) in Myr,
$Z$ the metallicity in solar units, and $t_{sfr}$ the $e$-folding timescale of star formation in Myr
as a function of $\tau_{in}/(\tau_{in} + \tau_{sc})$. 
The left panel in the bottom row plots the best-fit SED along with the unreddened SED and the right 
panel the contour map of $\chi_{\nu}^2 - \chi_{\nu}^2(min)$ where $\chi_{\nu}^2(min)$ is 
the smallest $\chi_{\nu}^2$ among all the models.

The trends of the results are:(1) the SN dust curves generally have reducing $\chi^2_{\nu}$ as 
$\tau_{in}/(\tau_{in} + \tau_{sc})$ is approaching to zero (more dominated by screen dust) 
and $N$ approaching to $\infty$ (more uniform screen); 
(2) Some galaxies have a small  $\chi^2_{\nu}$ in a clumpy foreground 
screen geometry, especially, galaxies Nos. 5 and 12 have the minimum $\chi^2_{\nu}$ 
at $N$ of 10 and 3,respectively; 
(3) The MW and pseudo MW curves sometimes have reducing $\chi^2_{\nu}$ as a geometry gets 
more dominated by internal dust at $\tau_{in}/(\tau_{in} + \tau_{sc})$ = 0.8 - 0.9.  
The data point around rest 4\micron\  in the SED plots is the IRAC 8.0 \micron\ data, and 
shows an 
excess from the model for most of the galaxies. As discussed in section 3.3, the data of this band 
were not used in the analysis, because dust emission may be significant in this band. 
The SED plots in Figure \ref{f_result} indeed suggest dust emission around 4\micron.

Table \ref{t_chi2} compares the smallest $\chi_{\nu}^2$ of the respective extinction curves 
along with the dust geometry, the scattering property, and the total extinction including 
internal and screen dust.There are two obvious features:
(1) the optical depth ($A_{0.3\micron}$) in the geometry dominated by internal 
dust is several times greater than that in the foreground screen geometry; 
(2) the MW curve generally has the smallest $\chi_{\nu}^2$ in the composite geometry dominated 
by internal dust, whereas the SN curves do 
in the foreground screen geometry, and the pseudo MW dust model is in midway.  Finding of
an higher optical depth for the internal absorption is an expected and obvious result, 
because the screen geometry always maximizes the extinction.

The contour plots in Figure \ref{f_result} appear to show the 
age($t$)/extinction($A_{0.3um}$) degeneracy, suggesting that $t$ and $t_{sfr}$ given 
in Table \ref{t_chi2} are not very accurate. 
The uncertainties in the age determination were propagated 
to the uncertainty to determine the stellar mass of galaxies, because the luminosity to mass ratio 
strongly depends on the age of stellar populations. This degeneracy can be resolved by adding 
data from 100\micron\ to the submillimeter, greatly reducing the present uncertainty in $L_{IR}$.

Table \ref{t_fit_para} summarizes the best-fit models.  Out of our 12 galaxies, 10 (83\%) have 
a pure foreground dust screen, and 9 of the 10 galaxies have a uniform foreground screen. 
8 (67\%) of the 12 galaxies have the best-fit model with one of the SN dust curves, all of which are 
associated with a pure foreground dust screen. 
Internal dust is only significant in two galaxies, all of which have the best-fit models with
the MW or pseudo MW extinction curve.
 Because the extinction curves analyzed here can be ordered from the steepest to the flattest as
the MW, pseudo MW, Calzetti, 3-40 $M_{\sun}$ SN, and 3-30 $M_{\sun}$ SN, the MW and pseudo MW curves 
are too steep to agree with a pure foreground screen, 
and need to be associated with a geometry dominated by internal dust or a clumpy foreground 
screen to make the attenuation curve flat or gray. This requires a few times more dust mass 
than that in a pure uniform foreground screen. Our results indicate 
that the 8-40 $M_{\sun}$ SN curve with a uniform foreground screen is the best dust model 
to account for $z \sim 1$ ULIRGs. 

\begin{table*}
	\begin{minipage}{170mm}
	\centering
	\caption{Properties of the best-fit models.}
	\setlength{\tabcolsep}{5pt}
	\begin{tabular}{cccllccccccccc}
	\hline
No. & & $\tau_{in}/\tau_{in+sc}$ $(N)$  & Dust model  & Scattering & $\chi^2_\nu$ & 
$Z$ & $t$ & $t_{sfr}$ & $A_{0.3\micron}$ & $M_{in}$ & $M_{sc}$ & $M_*$ & $M_d$/SN\\
(1) & & (2)  & (3) & (4) & (5) & (6) & (7) & (8) & (9) & (10) & (11) & (12) & (13)\\
	\hline
	\hline
1  & & 0.0 ($\infty$) & SN40 & No      & 0.79 & 0.2 & 1500 & 10000 & 4.3 & - & 1.6 & 27.5 & 0.0079\\
2  & & 0.0 ($\infty$) & SN40 & No      & 1.61 & 1.0 & 800 & 630 & 3.4 & - & 1.9 & 6.5  & 0.031\\
3  & & 0.0 ($\infty$) & SN40 & No      & 1.62 & 1.0 & 60 & 10 & 6.0 & - & 1.0 & 1.4  & 0.16\\
4  & & 0.0 ($\infty$) & pMW  & Isotrop & 0.86 & 1.0 & 100 & 40 & 5.9 & - & 1.3 & 1.2  &0.19 \\
5  & & 0.0 (10)       & SN40 & No      & 0.80 & 0.4 & 60 & 20 & 4.8 & - & 2.8 & 1.9  & 0.25\\
6  & & 0.0 ($\infty$) & SN40 & Isotrop & 1.37 & 1.0 & 1500 & 320 & 4.9 & - & 1.7 & 11.3  & 0.017\\
7  & & 0.0 ($\infty$) & SN40 & No      & 2.35 & 0.4 & 40 & 10 & 3.6 & - & 2.5 & 0.6  & 0.38\\
8  & & 0.0 ($\infty$) & SN40 & No      & 0.98 & 1.0 & 150 & 40 & 3.6 & - & 1.2 & 1.5  & 0.085\\
9  & & 0.9 (--)       & pMW  & Isotrop & 1.02 & 1.0 & 150 & 160 & 21.3 & 2.4 & 0.8 & 1.5  & 0.23\\
10 & & 0.0 ($\infty$) & Calzetti & No  & 0.75 & 0.4 & 60 & 10 & 3.6 & - & 1.0 & 1.0  & 0.12\\
11 & & 0.0 ($\infty$) & SN30 & No      & 3.56 & 1.0 & 630 & 160 & 5.4 & - & 0.9 & 3.6  & 0.058\\
12 & & 0.4 (3) &  MW  & Isotrop        & 0.84 & 0.4 & 2000 & 630 & 20.0 & 0.7 & 2.2 & 42.0  & 0.0098\\
   & & 0.4 (3) & pMW  & Isotrop        & 0.84 & 0.4 & 2000 & 630 & 20.0 & 0.7 & 2.2 & 42.0  & 0.0098\\
	\hline
	\end{tabular}
	\label{t_fit_para}
	\begin{flushleft}
	\medskip
\textbf{Note.} - 
Columns 1-5 are defined in Table \ref{t_chi2}.\\
Columns 6-9: Metallicity($Z$) in solar, age($t$) in Myr, e-folding timescale($t_{sfr}$) in Myr, 
and extinction at 0.3\micron $(A_{0.3\micron})$ in magnitude, respectively.\\
Columns 10-11: Mass of internal dust ($M_{in}$) and screen dust ($M_{sc}$), respectively,  
in units $10^8$ solor mass. \\
Column 12: Stellar mass ($M_*$) in units $10^{11} M_{\sun}$ that is estimated as $L_{IR}/(L/M_*)$, 
		where $L/M_*$ is the luminosity to mass ratio for the stellar population models.\\
Column 13: Dust mass($M_d$) per SN where ($M_d$) in solar units. The number of SNs per galaxy 
           is derived from the IMF of the model of star formation history and integration of all stars 
           with $\gtrsim$ 8 M$_{\sun}$.
	\end{flushleft}
	\end{minipage}
\end{table*}

\section{Discussion}

\citet{Gordon1997} applied models of stars and dust to the SEDs of 30 local starburst galaxies 
using the MW and SMC extinction curves, and found that the dust has an extinction curve 
lacking a 2175 \AA\ feature with a slope intermediate between the MW and SMC curves. 
The dust geometry has a pure foreground screen with clumps. \citet{Gaskell2004} found 
indications of a very flat extinction curve in the ultra-violet(UV) from an analysis of 1,090 AGNs. 
Large samples of SDSS quasars indicate that the extinction law in quasars at $z < 2.2$ is 
described by SMC reddening but not by reddening in the LMC and MW 
(\citealt{Richards2003}; \citealt{Hopkins2004}). \citet{Gallerani2010} proposed that 
 reddening of quasars at $z = 3.9 - 6.4$ has a tendency to be flat at 
$\lambda \lesssim$ 2000 \AA. \citet{Maiolino2004} found that the BALQSO at $z$=6.2, 
requires a flat extinction curve 
which is quite different from those observed in the MW, LMC and SMC, 
but in excellent agreement with the synthetic SN curves \citep{Todini2001}. 
The flat extinction law 
that we found in $z \sim 1$ ULIRGs are probably incompatible with the steep SMC-like extinction law 
reported for $z < 2.2$ AGNs, and  
in agreement with the flat extinction law found in high redshift quasars.

It should be reminded that the SN-dust models generally predict
flat extinction curves but there are intrinsic differences among
the models. \citet{Maiolino2004} adopted the SN-dust model by
\citet{Todini2001} to derive the extinction curves expected
in high-z quasars. The calculated curves show a flat feature at
UV wavelengths as a result of the main contribution of amorphous
carbon grains with radii around 300 \AA, added by the minor
contributions of small silicate and Fe$_3$O$_4$ grains with radii around 20 \AA.
\citet{Bianchi2007} revisited the dust formation model by
\citet{Todini2001} and also followed the evolution of the dust
grains from the time of formation to their survival through the
passage of the reverse shock. They showed that the reverse shock
predominantly destroys small silicate and Fe$_3$O$_4$ grains, and the
resultant extinction curves become flatter, being dominated by
carbon grains whose extinction coefficients are relatively insensitive
to wavelengths.
On the other hand, \citet{Hirashita2005} calculated the extinction
curves using SN II-dust model by \citet{Nozawa2003}. The derived
extinction properties more or less depend on the SN progenitor
mass, but are quite flat as a result of the dominant contribution of
large ($\gtrsim$ 0.1 \micron) amorphous carbon or Si grains.
As discussed in Section 3.1.2, the reverse shock efficiently destroys
small grains \citep{Nozawa2007}, making the extinction curves much
flatter than the original curves (see \citealt{Hirashita2008}).
Then large grains that survive the destruction through the reverse
shock are shattered in the WIM, producing a large number of small
grains. This finally leads to the extinction curves similar to the
original SN extinction curves \citep{Hirashita2010}. 
The growth of dust grains through accretion in the ISM would also be efficient in 
infrared galaxies(\citealt{Dwek1998};\citealt{Draine2009};\citealt{Michalowski2010}), 
and also cause flat extinction curves.

As shown in Table \ref{t_fit_para}, our $z \sim 1$ ULIRGs have an age of mostly $t \leq 650$ Myr 
with four exceptions (i.e., $t$ = 800, 1,500, 1,500, 2,000 Myr), 
a stellar metallicity of $z = 0.2 - 1.0$ solar,
and a dust mass of $M_d \sim$ $10^8 M_{\sun}$. 
\citet{Michalowski2010} suggested that AGB stars are not 
sufficient enough to form dust at such an early epoch of galaxy evolution history, 
whereas SN-condensed dust followed by grain growth in the ISM is important to account for 
the inferred dust mass. It is noted that our analysis cannot rule out a possible contribution 
from AGB stars, because some observations (\citealt{Groenewegen1997};\citealt{Gauger1999}) 
suggest that large-sized dust grains are formed in AGB stars 
and larger grains have flat extinction curves.
The dust mass per SN in our $z \sim 1$ ULIRGs mostly ranges from 0.01 $M_{\sun}/SN$ 
to 0.3 $M_{\sun}/SN$. 
This should be compared with theoretical prediction of the dust yield per SN, 
which is 0.01 - 0.1 $M_{\sun}/SN$ with dust destruction and 
0.1 - 1 $M_{\sun}/SN$ without dust destruction. If the ISM process of dust growth is important, 
the expected yield will be more than these. However there would be a possible bias effect on 
the dust mass per SN inferred from the observations, because a significant correlation 
between the age $t$ and $M_d/SN$ can be seen in Table \ref{t_fit_para}. This correlation seems 
spurious and simply reflects the uncertainty of determining $t$ caused by 
the $t$ and $A_{0.3\micron}$ degeneracy, 
because the mass-to-luminosity ratio strongly depends on $t$.

Our results indicate that the dust geometry of a pure foreground dust screen generally 
explains the data of the $z \sim 1$ ULIRGs better than that dominated by internal dust. 
Color-color plots of optical and UV data of local starbursts were compared with various 
dust distribution models by \citet{Gordon1997}, who found that the observed spread cannot be 
explained by the internal dust geometry, but the foreground screen 
geometry can explain it. 
\citet{Calzetti1994} also showed the failure of the internal dust geometry to 
explain the spread of data of local starbursts, suggesting the foreground screen geometry 
with a flat extinction curve (i.e., Calzetti curve). 
\citet{Meurer1995} showed that a strong correlation between the 
far-IR excess and the UV spectral slope for local starbursts can be explained by dust 
in a foreground dust screen geometry with the Calzetti curve, but not by an internal dust geometry. 
They attributed the dust in a foreground screen to the Galactic winds frequently observed 
in starbursts that will sweep out any diffuse ISM from the starburst sites on a timescale 
of a few Myr \citep{Heckman1990}. The foreground dust screen geometry we are looking at
in $z \sim 1$ ULIRGs may also result from the dust that was swept by the Galactic winds.

\section{Conclusions}

In order to study the extinction law in galaxies (ULIRGs) at $z \sim 1$, we have analyzed the 
multi-wavelength photometric and spectroscopic data of 12 ULIRGs which are archived or 
published in the literature; 10 galaxies come from the COSMOS survey catalogs, and the others are 
SST J1604+4304 and HR10.  

The data are compared with models of stars and dust. We applied 
five extinction curves, namely, the Milky Way (MW), the pseudo MW which is MW-like 
without the 2175 \AA\ feature, the Calzetti, and two SN dust curves, 
by combining with various dust distributions, namely, the foreground screen geometry, 
the internal dust geometry, and the composite geometry with a combination of 
dust screen and internal dust. Several scattering properties are used: 
no scattering, isotropic scattering, and forward scattering. In case of no scattering 
the effect of dust to convey photons into the line of sight is ignored, while this effect 
is included in isotropic scattering and forward scattering.

The two SN extinction curves that we applied here were 
calculated by \citet{Hirashita2005} based on the grain properties of SN-condensed 
dust derived by \citet{Nozawa2003}. These SN extinction curves were calculated assuming the 
unmixed ejecta of SNe II for the Salpeter IMF in a progenitor mass range of 8 - 30 or 
8 - 40 $M_{\sun}$ and do not include the effects of the reverse shock in the ejecta and 
interaction between SN dust grains and the ambient ISM. 

Employing a minimum  $\chi^2$ method, we find that the foreground dust screen geometry, especially 
combined with the 8 - 40 $M_{\sun}$ curve,  
provides a good approximation to the real dust geometry, whereas 
internal dust is significant only for 2 galaxies. 
The SN extinction curves, which are flatter than the other extinction curves, reproduce 
the data of 8(67\%) galaxies better than the other curves do. 
Our $z \sim 1$ ULIRGs require an age of mostly $t \leq 650$ Myr, 
a stellar metallicity of $z = 0.2 - 1.0$ solar,
and a dust mass of $M_d \sim$ $10^8 M_{\sun}$. 
Considering these physical features, we conclude SN-origin dust is the most plausible 
to account for the vast amount 
of dust masses and the flat slope of the observed extinction law.
The inferred dust mass per SN ranges from 0.01 to 0.4 M$_{\sun}$/SN.

\section{Online Materials}

The full version of the paper is for online publication only, where the full version of 
Figure \ref{f_result} is available.




\section*{Acknowledgments}

We thank H. Hirashita for helpful comments and suggestions and Brice M\'{e}nard for his critical 
reading of the final draft. Simon Bianchi provided us with extremely constructive comments and 
criticism which improved this work very much.
This work has been supported in part by Grants-in-Aid for Scientific research
(20340038, 22111503) and Specially Promoted Research (20001003) from JPSP.

\nocite{*}
\bibliographystyle{mn2e}
\bibliography{ref}

\clearpage

\label{lastpage}

\end{document}